\pdfoutput=1
\documentclass[10pt,twocolumn]{article}
\usepackage{amsmath,amssymb}
\usepackage{graphicx}
\usepackage{color}
\usepackage{textcomp} 
\usepackage{listings}
\usepackage[backend=biber,bibencoding=utf8, style=numeric,sorting=none]{biblatex}
\usepackage[top=2cm, bottom=2.5cm, left=2cm, right =2cm]{geometry}
\bibliography{references}
\usepackage{breqn}
\usepackage[affil-it]{authblk}
\usepackage{abstract}
\usepackage{tabularx}
\usepackage{hyperref}
\hypersetup{
    bookmarks=true,         
    unicode=false,          
    pdftoolbar=true,        
    pdfmenubar=true,        
    pdffitwindow=false,     
    pdfstartview={FitH},    
    pdfauthor={Anshuman Baruah},     
    colorlinks=true,       
    linkcolor=blue,          
    citecolor=red,        
}

\makeatletter
\newbox\abstract@box
\renewenvironment{abstract}
  {\global\setbox\abstract@box=\vbox\bgroup
     \hsize=\textwidth\linewidth=\textwidth
    \small
    \begin{center}%
    {\bfseries \abstractname\vspace{-.5em}\vspace{\z@}}%
    \end{center}%
    \quotation}
  {\endquotation\egroup}
\expandafter\def\expandafter\@maketitle\expandafter{\@maketitle
  \ifvoid\abstract@box\else\unvbox\abstract@box\if@twocolumn\vskip1.5em\fi\fi}
\makeatother
\begin{document}
\definecolor{dkgreen}{rgb}{0,0.6,0}
\definecolor{gray}{rgb}{0.5,0.5,0.5}
\definecolor{mauve}{rgb}{0.58,0,0.82}

\lstset{frame=tb,
  	language=Matlab,
  	aboveskip=3mm,
  	belowskip=3mm,
  	showstringspaces=false,
  	columns=flexible,
  	basicstyle={\small\ttfamily},
  	numbers=none,
  	numberstyle=\tiny\color{gray},
 	keywordstyle=\color{blue},
	commentstyle=\color{dkgreen},
  	stringstyle=\color{mauve},
  	breaklines=true,
  	breakatwhitespace=true
  	tabsize=3
}


\title{Traversable Wormholes in Higher Dimensional Theories of Gravity}
\author{Anshuman Baruah \thanks{E-mail: \textsf{anshuman.baruah@aus.ac.in}} \textit{and} Atri Deshamukhya} 
\affil{Department of Physics, Assam University, Silchar, Assam, India-788011}
\date{26 March 2019}

\begin{abstract}
Wormhole solutions in classical General Relativity are unstable and hence non traversable. Morris and Thorne discovered a traversable wormhole solution that required the energy momentum tensor of matter sources to violate various energy conditions and are out of the purview of the standard model of particle physics. The search for traversable wormhole solutions in  Modified Theories of Gravity has been of significant interest in the decades after Morris and Thorne first published their results as such violations may be avoided in such theories. This work comprehensively reviews traversable wormhole solutions in modified theories of gravity with extra dimensions that satisfy the various energy conditions with an in depth look at the matter sources and the various constraints on the parameters of the theory to make the energy momentum of the matter sources respect the energy conditions.
\end{abstract}
\maketitle
\section{Introduction}
\label{intro}
\begin{figure}
  \begin{center}
    \includegraphics[width=0.4\textwidth, height=0.4\textwidth, keepaspectratio]{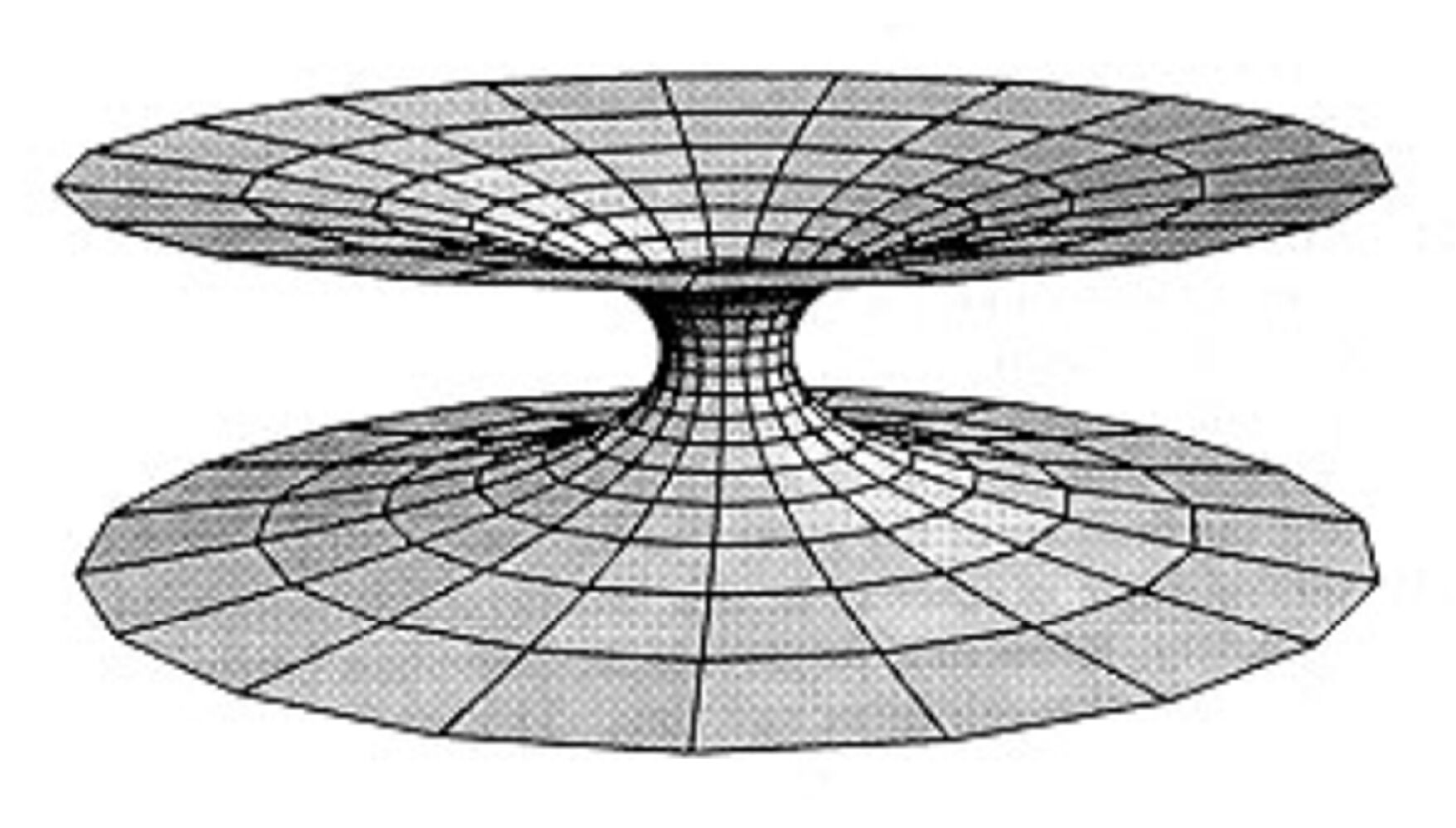}
\caption{Embedding diagram for MT wormhole which is a 2-surface ($\theta = \pi/2$, $t=constant$) embedded in 3D Eucledian space}
\label{figure1}
  \end{center}
  \end{figure}

Wormhole solutions to the Einstein's Field Equations (EFE) have been studied since the early days of the General Theory of Relativity (GTR). They are essentially shortcuts through space-time that can connect vast distances in our universe or possibly connect our universe to another. Visser defines wormholes as any compact region of space-time with a topologically simple boundary but a topologically non trivial interior \cite{Visser:1995cc}. The earliest examination of such a solution was done by Flamm in 1916 \cite{Flamm:1916}. Einstein and Rosen in their seminal 1935 work \cite{Einstein:1935tc} tried to develop a unified theory of electromagnetism and gravity and therein were the first to interpret such a solution as two asymptotically flat regions of space-time connected by a tube or `bridge'. Their solution however was geodesically incomplete due to the presence of a physical singularity. It was also shown in \cite{Guendelman:2016bwj} that the Einstein-Rose bridge is not equivalent to the concept of a dynamic non-traversable wormhole solution. Ellis was the first to describe a geodesically complete static, spherically symmetric, horizonless space-time manifold connecting two asymptotically flat regions of space-time \cite{Ellis:1973yv}. These solutions are referred to in literature as Einstein-Rosen Bridges or Lorentzian wormholes. Wheeler for the first time used the term wormhole in \cite{Wheeler:1957} to desrcibe such solutions. Such wormholes are non-traversable as they are unstable and their throats `pinch off' far too quickly for light that falls in one exterior region to emerge in the other exterior region \cite{PhysRev.128.919}. Morris and Thorne in their seminal  1987 paper described the first traversable wormhole metric \cite{Morris:1988cz}. Their line element in Schwarzschild coordinates is given by:  
\begin{equation}
ds^{2}=e^{2\Phi(r)}dt^{2}-\frac{dr^{2}}{1-\frac{b(r)}{r}}-r^{2}[d\theta^{2}+\sin^{2}\theta d\varphi^{2}]
\label{eq1}
\end{equation}
Where, $\Phi(r)$ determines the gravitational redshift and is called the redshift function. $b(r)$ determines the spatial shape of the wormhole and is called the wormhole shape function. In the following discussion, we will see following the treatment of Morris and Thorne, how this metric is interpreted as a wormhole and how the constraints on the metric functions impose via the EFEs constraints on the matter threading the wormhole. To interpret that the metric \eqref{eq1} indeed describes a wormhole, one takes the help of geometrical embedding diagrams. It is useful to consider a three dimensional space at a fixed time with spherical symmetry which can be obtained by setting $t=constant$ and $\theta=\pi/2$ in \eqref{eq1}. We wish to visualize this equatorial ($\theta=\pi/2$) slice as removed from the space-time of \eqref{eq1} and embedded in Euclidean space. We consider cylindrical coordinates ($z,r,\varphi$) in this embedding Euclidean space which is axially symmetric and thus can be defined by some function $z=z(r)$. The line element on that surface will be:
\begin{equation}
ds^{2}=\left[1+\left(\frac{dz}{dr}\right)^{2}\right]dr^{2}+
r^{2} d\varphi^{2}
\label{eq2}
\end{equation} 
This surface can be interpreted as the equatorial slice of the solution \eqref{eq1} if we identify the coordinates $(r,\varphi)$ of the embedding space and the wormhole space-time as the same and with the requirement:
\begin{equation}
\frac{dz}{dr}=\pm \left(\frac{r}{b(r)}-1\right)^{-\frac{1}{2}}
\label{eq3}
\end{equation}
This surface has been visualized for arbitrary values of the parameters in Figure \ref{figure1}. Equation \eqref{eq3} shows how the function $b(r)$ shapes the wormhole's spatial geometry, and hence the name shape function. \\
The field equations derived from the metric \eqref{eq1} can be suitably rearranged to obtain expressions for the mass energy density $\rho$, radial tension per unit area $\tau$ and the lateral pressure $p$. These expressions are: 
\begin{eqnarray}
\rho=b'/[8\pi G c^{-2}r^{2}] \\
\tau=[b/r-2(r-b)\Phi ']/[8\pi G c^{-4}r^{2}] \\
p=(r/2)[(\rho c^{2} - \tau)\Phi'-\tau ']-\tau
\label{eq4}
\end{eqnarray}
Here, prime denotes derivative with respect to the $r$ coordinate. Every wormhole has some minimum radius $r=b_{o}$ at which the expression \eqref{eq3} is divergent, meaning $b(r)=r$ and that the embedding surface becomes vertical. This region is called the wormhole throat and the minimum radius $r=b_{o}=b$ is called the throat radius. Another dimensionless function is defined to study the tension at the throat region as $\varsigma \equiv \frac{\tau-\rho c^{2}}{|\rho c^{2}|}$. The usefulness of $\varsigma$ will be apparent from the upcoming discussion. As is apparent from \eqref{eq1}, the $r$ coordinate is singular at the throat and hence the proper radial distance:
\begin{equation}
l(r)=\pm {\int_{b_{o}}}^{r} \frac{dr}{[1-b(r)]^{\frac{1}{2}}}
\label{eq5}
\end{equation}
must be well behaved or finite all throughout the space-time. This implies $1-\frac{b}{r} \geq 0$ throughout the space-time. Far from the throat, the space-time becomes asymptotically flat in both radial directions. This requirement implies that the embedding surface flares out at the throat. Mathematically, this means that at the throat:
\begin{equation}
\frac{d^{2}r}{dz^{2}}=\frac{b-b'r}{	2b^{2}} >0
\label{eq6}
\end{equation}
Using Equation \eqref{eq6} and the definition of $\varsigma$,we can rewrite Equation \eqref{eq6} as:
\begin{equation}
\varsigma_{o}=\frac{\tau_{o}-\rho_{o} c^{2}}{|\rho_{o} c^{2}|}>0
\label{eq7}
\end{equation}
Here, the subscript `$o$' denotes values of the parameters at the throat. Equations \eqref{eq6} and \eqref{eq7} are known as the \textit{flaring out conditions} as obtained by Morris and Thorne. \\
The condition for traversability demands that the throat possess no horizon. Horizons in spherically symmetric space-times are identified by physically non singular surfaces at $g_{00}=-e^{2\Phi}\rightarrow 0$. This implies the constraint that $\Phi(r)$ must be everywhere finite. These constraints on the metric functions impose in turn via the EFEs, constraints on the mass-energy density, radial tension and the lateral pressure of the matter that threads the geometry. In the context of GTR and other field theories, the \textit{energy conditions} are sets of inequalities or relations that the matter energy momentum tensor (EMT) is required to respect so that the energy density of the matter fields is measured to be positive by any observer traversing a time-like curve. The physical and effective interpretation of some relevant energy conditions in GTR has been summarized in Table \ref{table1}, for a diagonal EMT with the timelike-timelike component interpreted as the ordinary mass-energy density and the three space-like components as the three pressures.
\begin{table}[h!]
\centering
\caption{Energy Conditions in GTR} \label{table1}
\begin{tabularx}{\columnwidth}{@{\extracolsep{\fill}}|X|l|p{1.5cm}|}

\hline
{\centering Energy Condition} & {\centering Physical} & {\centering Effective}  \\ 
\hline 
Null (NEC) & $T_{\mu \nu}k^{\mu} k^{\nu} \geq 0$ & $\rho+p \geq 0$ \\
Weak (WEC) & $T_{\mu \nu}X^{\mu} X^{\nu} \geq 0$ & $\rho \geq 0$ \\
Strong (SEC) & $\left(T_{\mu \nu}-\frac{1}{2}T g_{\mu \nu}\right)  X^{\mu} X^{\nu} \geq 0$ & $\rho+3p \geq 0 $ \\
\hline
\end{tabularx}
\end{table}

The costraints on the metric parameters of the Morris-Thorne wormhole discussed above, demands an abnormally high tension at the throat where it is required that $\tau_{0}>{\rho_{0}c}^{2}$. This means an observer moving through the throat with arbitrarily high velocities, observes a negative density of mass energy which is unphysical. This is the violation of the Weak Energy Condition (WEC) $\rho=T_{\mu \nu} X^{\mu} X^{\nu}  \geq 0 $, where X is any time-like vector field and $T_{\mu \nu}$ is the Energy Momentum Tensor (EMT) of the matter source. A more fundamental (weakest) energy condition is the Null Energy Condition (NEC)
  given by $T_{\mu \nu}k^{\mu} k^{\nu} \geq 0$ ($k$ is any null-like vector field)which is also violated. Such energy condition violating matter is referred to as `exotic' matter. Supplemented by additional constraints, the NEC/WEC lead to the other pointwise and averaged energy conditions such as the Srong and the Dominant and hence a violation of the WEC usually implies the violation of all the energy conditions in General Relativity. For a detailed review of energy conditions refer to \cite{Curiel:2014zba}. Though Quantum Field Theory allows the existence of arbitrarily negative energy states such as in the Casimir Effect, no known material or particle/field in the Standard Model has this property $\tau > \rho$. Moreover, the existence of such material on a macroscopic scale is highly debatable as of now. Many solutions around this problem have been devised such as limiting the amount of exotic matter to arbitrarily small regions, demanding finite radial cut off of the stress energy etc. Traversable wormhole geometries with minimal use of exotic material have also been constructed in thin-shell \cite{Visser-1989}, rotating \cite{PhysRevD.58.024014} and dynamic systems \cite{PhysRevLett.81.746, PhysRevD.49.862}. \\
  Thus, spherically symmetric traversable wormhole solutions in classical general relativity require exotic matter to exist. The same is true for non static, non spherical traversable wormholes \cite{Morris:1988cz}. However, in modified theories of gravity, as the structure of the field equations are different from those in classical General Relativity, a violation of the Null Convergence Condition $R_{\mu \nu}k^{\mu} k^{\nu} \geq 0$ (which all Lorentzian wormholes violate) does not necessarily lead to a violation of the Null Energy Condition (NEC) and hence spherically symmetric wormhole solutions have been extensively studied in theories such as $f(R)$ Gravity \cite{PhysRevD.80.104012, PhysRevD.87.067504}. Modified theories of gravity with extra physical dimensions (compactified or large) are very useful in unification and also in other aspects of particle physics, cosmology and astrophysics. Traversable wormhole solutions can be obtained in many such theories with matter sources that do not necessarily violate the WEC or the NEC. This work reviews and generalizes the most important conditions in various higher dimensional gravity theories which are imposed on the matter energy momentum tensor to make it respect the energy conditions. The violations however may arise from the higher order curvature terms or from effective energy momentum tensor components as we will see in detail in the coming sections. \\
This review is organized in the following way: In Section \ref{intro}, we have begun by introducing the traversable wormhole solution of Morris and Thorne. In Section \ref{whinmtg} we briefly review wormhole solutions in some extended/Modified theories of gravity with a look at the matter content and energy condition violations in each theory. In Section \ref{whhd} we present a general introduction of Modified theories of gravity with extra physical dimensions and then proceed to review in detail some wormhole solutions widely studied in the context of Modified theories of gravity with extra physical dimensions and finally in Section \ref{cnd} we conclude with discussions and remarks. We hereafter use natural system of units where the constants such as $G,c$ etc. have been set to unity. the metric signature under consideration in the rest of the paper is $(-,+,+,+)$.
\section{Wormholes in Modified theories of gravity}
\label{whinmtg}
Einstein's GTR, though well verified in the solar system fails to provide explanations for observed cosmological phenomena such as Dark Matter and Dark Energy. After the advent of Quantum Mechanics it was also necessary to formulate a quantum theory of gravitation in order to unite it with the other fundamental forces. The earliest of such attempts to develop a unified theory of gravitation and electromagnetism can be traced back to the work of Einstein \& Rosen and Kaluza \& Klein \cite{Kaluza:1921tu, Einstein:1935tc}. In GTR, the gravitational field is assumed to be mediated by a second rank tensor field. This is entirely geometric in nature. In the context of Quantum Field Theory , gravitation is mediated by the massless spin-2 \textit{graviton}. Apart from the coupling of the matter fields of GTR, it is plausible to assume couplings of other fields in the field equations of gravitation; the effects of which must in some way vanish at length scales comparable to the solar system where GTR has been well tested. One fundamental approach is to consider the weak coupling of a scalar field to the gravitational sector of the field equations. These are known as \textit{Scalar Tensor Theories}. Another approach is to take into consideration higher dimensions. \\
In this section, we present an overview from literature of wormhole solutions widely studied in modified theories of gravity without considering extra dimensional effects and in Section \ref{whhd} we discuss wormhole solutions in higher dimensional gravity theories. A detailed account of modified theories of gravity is out of the scope of this article and we refer the reader to \cite{CLIFTON20121} for a detailed review of the subject.\\
In classical GTR, as mentioned before, wormhole solutions violate the null convergence condition and in return, the NEC. This actually is inherent from the fact that a wormhole must defocus a set of congruent null geodesics. Positivity of $T_{\mu \nu}k^{\mu} k^{\nu}$ ensures that the geodesic congruences focus within a finite value of the affine parameter. However, in Modified theories of gravity, the Einstein's field equations are written with an effective energy momentum tensor $T_{\mu \nu}^{eff}$. 
\begin{equation}
G_{\mu \nu}=R_{\mu \nu}-\frac{1}{2}g_{\mu \nu}R=T_{\mu \nu}^{eff}
\label{eq8}
\end{equation}
This effective EMT contains the matter energy momentum tensor $T_{\mu \nu}$ and the higher order curvature terms arising in different such theories. In such cases, it is possible to impose that while the respective generalized NEC $T_{\mu \nu}^{eff}k^{\mu} k^{\nu} \geq 0$ is violated, the matter EMT $T_{\mu \nu}$ satisfies the NEC. An example is $f(R)$ gravity where $f(R)$ is some function of the Ricci Scalar and the considerations on $f(R)$ gives a family of different theories. GTR is received back when $f(R)=R$. The action of this theory is defined as:
\begin{equation}
S= \int d^{4x} \sqrt{-g} \left[\frac{1}{2 \kappa^{2}}f(R)+{\mathcal{L}}_{m} (g_{\mu \nu}, \psi)\right]
\label{eq9}
\end{equation}
Here, $\psi$ collectively denotes all the mater fields and $\kappa$ is the gravitational coupling constant. The condition for the violation of the generalized WEC in $f(R)$ theories is given by \cite{PhysRevD.80.104012, PhysRevD.87.067504}:
\begin{equation}
\frac{1}{f_{r}} T_{\mu \nu} k^{\mu} k^{\nu} < -\frac{1}{\kappa^{2} f_{r}} k^{\mu} k^{\nu} \nabla_{\mu} \nabla_{\nu} f_{R} 
\label{eq10}
\end{equation}
Here, $f_{r} \equiv df/dR$. Depending on the form of $f(R)$ taken into consideration, this inequality can be fulfilled even if the term $T_{\mu \nu} k^{\mu} k^{\nu}>0$. The constraint $T_{\mu \nu} k^{\mu} k^{\nu} \geq 0$ is imposed when $f(R)>0$. Similar results can be obtained in $f(R,{\mathcal{L}}_{m})$ which generalizes $f(R)$ theories with curvature-matter coupled models \cite{PhysRevD.75.104016}. \\ 
Another extended theory of gravitation that has been extensively studied is $f(R,T)$ gravity where $T$ is the trace of the energy momentum tensor \cite{Harko:2011kv}. The action of this theory has the form:
\begin{equation}
S=\int d^{4} x\sqrt{-g} [f(R,T)+ {\mathcal{L}}_{m}]
\label{eq11}
\end{equation}
It has been shown that static, spherically symmetric wormhole solutions with anisotropic matter sources that respect the NEC can be obtained in this theory for $f(R,T)=R+2f(T)$, $f(T)$ being some arbitrary function of the trace of the EMT \cite{Azizi:2012yv}. Other static, spherically symmetric wormhole solutions in this formalism have also been constructed in \cite{Zubair:2016cde, Sahoo:2017ual, Moraes:2017mir}. \\
Traversable wormhole solutions have also been constructed in hybrid metric Palatani gravity \cite{PhysRevD.85.084016}, that obey the NEC. The general form of the action for this theory is given by: 
\begin{equation}
S=\frac{1}{2 \kappa^{2}} \int d^{4} x \sqrt{-g} [R+f(\mathcal{R})]+\int d^{4} x \sqrt{-g}{\mathcal{L}}_{m}
\label{eq12}
\end{equation}
Here, $\mathcal{R}=g_{\mu \nu}{\mathcal{R}}_{\mu \nu}$ is the Palatani curvature. This theory can be suitably transformed into a scalar-tensor theory with the introduction of an auxiliary scalar field $\phi$, thus redefining the action as:
\begin{equation}
S=\frac{1}{2 \kappa^{2}} \int d^{4} x \sqrt{-g} [R+\phi \mathcal{R}-V(\phi)]+\int d^{4} x \sqrt{-g}{\mathcal{L}}_{m}
\label{eq13}
\end{equation}
The constraint on the matter source at the wormhole throat ($r=b=b_{o}$) in this formalism assuming that the matter stress energy satisfies the NEC and that $1+\phi>0$ is given by:
\begin{equation}
0<T_{\mu \nu}k^{\mu} k^{\nu}\vert_{r=b_{o}} <\frac{1}{\kappa}\left. \left[k^{\mu} k^{\nu} \nabla_{\mu} \nabla_{\nu} \phi-\frac{3}{2\phi}k^{\mu} k^{\nu}\partial_{\mu} \phi \partial_{\nu} \phi \right]\right\vert_{r=b_{o}}
\label{eq14}
\end{equation}
Two traversable wormhole solutions (in the Morris-Thorne background) have been obtained in \cite{Capozziello:2012hr} with suitable choices of the metric functions and the scalar field, the first of which is not asymptotically flat. Traversable wormhole solutions that obey the NEC everywhere in a generalized hybrid metric Palatani formalism have been obtained in \cite{Rosa:2018jwp}.\\
Asymptotically flat and anti-de Sitter (AdS) wormhole space-times without exotic matter have also been obtained in Einstein-Cartan Gravity \cite{Mehdizadeh:2017tcf}. The action for this theory is given by:
\begin{equation}
S= \int d^{4}x \sqrt{-g} \left\lbrace \frac{-1}{2\kappa}(\tilde{R}+2 \Lambda)+{\mathcal{L}}_{m} \right\rbrace
\label{eq15}
\end{equation}
Here, $\tilde{R}$ is the Ricci Scalar constructed from the asymmetric Christoffel connections of the Einstein-Cartan manifold. A detailed review of the Einstein-Cartan Theory can be found in \cite{Hehl:2007bn}. $\Lambda$ is the cosmological constant. Wormhole solutions in this theory have been studied with a Wyssenhoff fluid \cite{Weyssenhoff:1947iua, Boehmer:2006gd} as the modified source, the spin tensor of which is given by: 
\begin{equation}
{\tau^{\alpha}}_{\mu \nu}=S_{\mu \nu} u^{\alpha}
\label{eq16}
\end{equation}
$u^{\alpha}$ being the four velocity of the fluid element and the second rank anti-symmetric tensor $S_{\mu \nu}$ is known as the spin density tensor. At the macroscopic scale, the square of this tensor given by $S^{2}=\frac{1}{2}\left\langle S_{\mu \nu}S^{\mu \nu} \right\rangle$ has contribution to the total effective energy momentum tensor. \\
In the Morris-Thorne background, at the wormhole throat $(r={r_{0}})$:
\begin{equation}
\left. \rho+p_{r} \right\vert_{r=b_{0}} = \frac{S^{2}(b_{0}){{r_{0}}}^{2}+2[b'(b_{0})-1]}{{2b_{0}}{^{2}}}
\label{eq17}
\end{equation}
which shows that the NEC $\rho+p_{r} \geq 0$ is violated for $S^{2}(b_{0})=0$. Traversable wormhole solutions can be then constructed by assuming $\sqrt{2({b'}_{0}-1)}/S_{0} < {r_{0}}$ and by choosing specific forms of the metric functions. Here $'$ denotes derivative with respect to $r$. Traversable wormhole space-times in ECT that are asymptotically flat and AdS have also been obtained with canonical scalar fields as sources with minimal and non minimal coupling to gravity \cite{Bronnikov:2015pha}. Dynamic (evolving) wormhole solutions (asymptotically flat and AdS) have also been found in ECT in a cosmological (FRW) background \cite{Mehdizadeh:2017dhb}.
\subsection{Generalization}
\label{gene}
Following the treatment of \cite{PhysRevD.87.067504}, one may generalize the conditions for traversable wormhole solutions in Modified theories of gravity in the following way. The generalized field equations can be written as:
\begin{equation}
g_{1} (\Psi^{i})(G_{\mu \nu}+H_{\mu \nu})-g_{2} (\Psi^{j})T_{\mu \nu}=\kappa^{2} T_{\mu \nu}
\label{eq18}
\end{equation}
Where $g_{1}(\Psi^{i})$, $(i=1,2)$ are multiplicative factors that modify the geometrical sector of the field equations and $H_{\mu \nu}$ is an additional geometric term that includes the additional curvature terms arising in Modified theories of gravity. $g_{2}(\Psi^{j})$ gives the coupling of the curvature invariants $\Psi$ (such as scalar fields) with the matter stress energy. The effective stress-energy tensor is given by:
\begin{equation}
{T^{eff}}_{\mu \nu} \equiv \frac{1+\frac{g_{2}(\Psi^ {j})}{\kappa^ {2}}}{g_{1}(\Psi^ {i})} T_{\mu \nu}-H_{\mu \nu}/\kappa^ {2}
\label{eq19}
\end{equation}
In order for the matter stress energy to respect the NEC, it is necessary to impose \cite{PhysRevD.87.067504}:  
\begin{equation}
T_{\mu \nu}u^{\mu} u^{\nu}=\frac{g_{1}(\Psi^ {i})}{\kappa^{2} +g_{2}(\Psi^ {j})} (G_{\mu \nu}+H_{\mu \nu})u^{\mu} u^{\nu} \geq 0
\label{eq20}
\end{equation}
Using Lorentz transformations, it is possible to show that $T_{\mu \nu}u^{\mu} u^{\nu} \geq 0$; which means that the energy density is measured to be positive in all local frames of reference.
\section{Wormholes in higher dimensional gravity theories}
\label{whhd}
\subsection{Modified theories of gravity with extra dimensions}
\label{mtged}
Space-time in GTR is a (3+1) dimensional Riemannian manifold. Riemannian geometry is not restricted to just (3+1) dimensions and hence one has the mathematical ability to probe theories of gravitation in higher dimensions. Even physically, the best candidate for a quantum theory of gravity is perhaps \textit{Superstring Theory} \cite{SCHWARZ1982223} which has been constructed on a 10 dimensional manifold. The phenomenological problem one faces in this approach is that gravity behaves strictly like a (3+1) dimensional theory would predict at least until length scales comparable to the solar system. Various approaches to account for this problem exist which lead to a host of different theories. \textit{Compactification} is one such approach wherein the extra dimensions are compactified or curled up into unobservably small length (energy) scales. The Kaluza-Klein (KK) \cite{Kaluza:1921tu, Klein:1926fj} theory put forward by T. Kaluza and O. Klein is the earliest such approach. The KK theory is a unified theory of gravitation and electromagnetism on a five dimensional manifold where the extra fifth dimension is compactified. An alternate approach is known as the \textit{Braneworld Scenario}. In this formalism, the extra dimensions can be infinitely large. The standard model particles are restricted to a (3+1) dimensional hyper-surface referred to as a \textit{brane}, embedded in some higher dimensional space-time known as the \textit{bulk}. The warped braneworld picture of Randall and Sundrum \cite{PhysRevLett.83.3370, Randall:1999vf} is a model that describes the universe as a five dimensional bulk anti-deSitter space-time with a (3+1) dimensional TeV brane and an extra dimensional \textit{Planck brane}. The Planck brane can have a finite size (RS1 Model) or maybe placed infinitely far away (RS2 Model). The standard model particles are confined on the TeV brane and gravity can live on the Planck Brane. A detailed review of the RS model can be found in \cite{1126-6708-2001-04-021}; and \cite{CLIFTON20121} has a detailed account of higher dimensional gravity theories which some readers may find helpful. In the forthcoming section, we discuss wormhole solutions in literature that respect the energy conditions in higher dimensional gravity theories.  
\subsection{Kaluza-Klein Theory}
As discussed in the Section \ref{mtged}, the Kaluza-Klein (KK) theory \cite{Kaluza:1921tu, Klein:1926fj} is the earliest precurssor of String models. It is a five dimensional extension to General Relativity. The primary success of Kaluza's theory was to show that five dimensional gravity contains Einstein's General Relativity as well as Maxwell's theory of electromagnetism and hence it was possible to unify gravitation and electromagnetism on a five dimensional manifold. The field equations are derived using a five dimensional form of the Einstein-Hilbert action:
\begin{equation}
S= \int \hat{R}\sqrt{-\hat{g}}d^{4} xdy
\label{eq21}
\end{equation} 
Where $y=x^{4}$ is the fifth coordinate and $\hat{R}$ is the 5D Ricci scalar and $\hat{g}$ is the five dimensional metric. Kaluza initially hypothesized what is referred to in literature as the \textit{Cylinder Condition} which means that while deriving the field equations, one drops any terms containing derivatives with respect to the fifth coordinate. The reduced field equations in four dimensions is:
\begin{equation}
\begin{split}
G_{\mu \nu}= \frac{\phi^{2}}{2} \left(g_{\alpha \beta} F_{\mu \alpha}F_{\nu \beta}-\frac{1}{4}g_{\mu \nu}F_{\alpha \beta}F^{\alpha \beta} \right) \\ + \frac{1}{\phi} (\nabla_{\mu} \nabla_{\nu} \phi-g_{\mu \nu} \square \phi)
\label{eq22}
\end{split}
\end{equation}
Here, $F_{\alpha \beta}=\partial_{\alpha}A_{\beta}-\partial_{\beta}A_{\alpha}$ and the first term inside brackets on the RHS of Equation \eqref{eq22} is the electromagnetic stress energy tensor. $\phi$ is the massless Kaluza-Klein scalar field. Kaluza's initial hypothesis \cite{Kaluza:1921tu} was
from an entirely classical point of view. Klein, in 1926 gave the quantum interpretation of the theory proposing that the fifth dimension is compactified \cite{Klein:1926fj}. We refer the reader to \cite{Overduin:1998pn} for a detailed review of the Kaluza-Klein Theory. Static, spherically symmetric wormhole solutions in KK theory was first explored in \cite{Ala-1982, PhysRevD.44.1330}. In 1982, Chodos and Detweiler derived the most general time independent, spherically symmetric, static solution to the Einstein’s field equations in five dimensions \cite{Ala-1982}. Their solution, with suitable choice of metric parameters, is analogous to the Schwarzschild solution of GTR. In the case when the charge to mass ratio of the gravitating body $Q/M>1$, the metric describes a wormhole connecting two asymptotically flat regions of space-time containing closed time-like curves. Axisymmetric wormhole solutions were found by Cl{\'e}ment \cite{Clement1984, Clement1984-2}. \\
An interesting solution is the case of evolving (non static) wormhole solutions that are traversable and satisfy the energy conditions in a Kaluza-Klein universe obtained by Kar and Deshdeep \cite{PhysRevD.53.722}. The space-time under consideration has a topology $R\otimes W \otimes S^{D}$ where $D$ is the dimension, $R\otimes W$ represents the wormhole space-time and $S^D$ is an extra dimensional 2-sphere. The metric describing such a universe is given by:
\begin{equation}
ds^{2}=-dt^{2}+{a^{2}}_{1}(t) \left[\frac{dr^{2}}{1-\frac{b(r)}{r}}+r^{2}d{\Omega^{2}}_{2} \right]+{a^{2}}_{2}(t)d{\Omega^{2}}_{D}
\label{eq23}
\end{equation}
Here, $a_{1}(t)$ and $a_{2}(t)$ are scale factors associated with the wormhole and the compact $D$-sphere respectively and ${\Omega^{2}}_{D}$ is the metric on the compact $D$-sphere. It is a known result that wormhole solutions with an inflationary scale factor in the metric cannot exist for any finite interval of time without violating the WEC \cite{PhysRevD.47.1370}. With a compact dimension however, an inflationary wormhole can exist for a finite interval of time without violating the WEC. The energy condition inequalities $\rho \geq 0$, $\rho+\tau \geq 0$, $\rho+p \geq 0$, for this ansatz of a Kaluza-Klein metric are:
\begin{equation}
3\left(\frac{{\dot{a}}_{1}}{a_{1}}\right)^{2}+3D\frac{{\dot{a}}_{1} {\dot{a}}_{2}}{a_{1} a_{2}}+\frac{D(D-1)}{2} \left[ \left(\frac{{\dot{a}}_{2}}{a_{2}}\right)^{2} +\frac{1}{{a^{2}}_{2}}\right]+ \\ \frac{b'}{r^{2}{a^{2}}_{1}} \geq 0 \label{eq24a} 
\end{equation}
\begin{equation}
-2\frac{\ddot{a_{1}}}{a_{1}}+2\left(\frac{{\dot{a}}_{1}}{a_{1}}\right)^{2}+D\frac{{\dot{a}}_{1} {\dot{a}}_{2}}{a_{1} a_{2}}-D\frac{\ddot{a_{2}}}{a_{2}}+\frac{b'r-b}{r^{3} {a^{2}}_{1}} \geq 0 \label{eq24b} 
\end{equation}
\begin{equation}
-2\frac{\ddot{a_{1}}}{a_{1}}+2\left(\frac{{\dot{a}}_{1}}{a_{1}}\right)^{2}+D\frac{{\dot{a}}_{1} {\dot{a}}_{2}}{a_{1} a_{2}}-D\frac{\ddot{a_{2}}}{a_{2}}+\frac{b'r-b}{2r^{3} {a^{2}}_{1}} \geq 0 \label{eq24c} 
\end{equation}
\begin{equation}
-D(D-1)\frac{\ddot{a_{2}}}{a_{2}}-3\frac{\ddot{a_{1}}}{a_{1}}+3\frac{{\dot{a}}_{1} {\dot{a}}_{2}}{a_{1} a_{2}}-D(D-1)\left[ \left(\frac{{\dot{a}}_{2}}{a_{2}}\right)^{2} +\frac{1}{{a^{2}}_{2}}\right] \geq 0
\label{eq24d}
\end{equation}
Here, overdot and prime represent derivatives with respect to $t$ and $r$ respectively. These inequalities can be satisfied for $b(r)=b_{o}$ with a perfect fluid matter source and with $0\leq a^{2} \leq \frac{dD}{d+D-1}$ (d and D are the number of normal and extra dimensions respectively). It is also seen from Equations \eqref{eq24a}-\eqref{eq24d} that for $d=4$ and $D=2$, \eqref{eq24a}, \eqref{eq24c} and \eqref{eq24d} are satisfied whereas \eqref{eq24b} constraints the value of $b(r)$. Such wormholes can therefore survive inflation into the current era, but their sizes will have grown infinitely large. Similar results were also obtained for FRW and exponential inflation as well in \cite{PhysRevD.53.722}. Static, spherically symmetric wormhole like space-times in KK theory with electric and magnetic fields have been found to exist for $0 \leq H_{KK} < E_{KK}$, where $H_{KK}$ and $E_{KK}$ are the Kaluza-Klein magnetic and electric fields respectively \cite{Dzhunushaliev:1999aka}. \textit{Dyonic} (characterized by two charge parameters: electric and magnetic charge) wormhole solutions similar to the ones obtained by Chodos and Detweiler were disovered by Chen \cite{ Chen:2000yi} wherein it was seen that the metric behaved as a Lorentzian wormhole when the electric charge is greater than the magnetic.  The particle model of this kind of solutions in the context of supergravity, can be interpreted as gravitational flux tubes containing electric and magnetic fields \cite{ Dzhunushaliev:2004pp}. The cross sections of these tubes are of the Planck order and they can be infinitely long when $E=H$, $E$ and $H$ being the electric and magnetic fields respectively. These flux tubes were called $Delta-strings$.  Wormhole solutions in 5D KK theory have also been found with the coupling of a massless ghost scalar field \cite{ Dzhunushaliev:2013iia}. The solution can be interpreted as a monopole from the fact that after the compactification of the fifth dimension, the radial magnetic field lines pass from one asymptotic region to the other and an observer sees a magnetic monopole. 
\subsection{Einstein-Gauss-Bonnet (EGB) Theory}
\label{egbwh}
EGB theory is a generalization of GTR to include the Gauss-Bonnet term given by ${R^{2}}_{GB}=R^{2}-4R^{\mu \nu}R_{\mu \nu}+R^{\mu \nu \rho \sigma} R_{\mu \nu \rho \sigma}$. This term is a generalization of the Gauss-Bonnet theorem and is non-trivial only in (4+1) D or higher. Lorentzian wormhole solutions in this theory was first obtained in \cite{PhysRevD.46.2464}. The action for the theory is given by:
\begin{equation}
\begin{split}
S=\int d^{4}x \sqrt{-g}[R+\alpha(R^{2}-4R^{\mu \nu}R_{\mu \nu}+R^{\mu \nu \rho \sigma} R_{\mu \nu \rho \sigma})]\\+\int d^{4}x {\mathcal{L}}_{m}
\label{eq25}
\end{split}
\end{equation} 
Here, $\alpha$ is the coupling coefficient of the Gauss-Bonnet combination. The WEC is satisfied for the matter sources threading wormhole geometries in this theory when the coupling constant of the Gauss-Bonnet combination is taken to be negative and is required to satisfy a certain inequality dependent on the wormhole shape function, the modulus of the coupling constant, $r$ and the number of space-time dimensions. The metric under consideration is of the Morris-Thorne type and is given by: 
\begin{equation}
ds^{2}=e^{2\Phi(r)}dt^{2}-\frac{dr^{2}}{1-\frac{b(r)}{r}}-r^{2}d{\Omega^{2}}_{D-2}
\label{eq26}
\end{equation}
Here, $d{\Omega^{2}}_{D-2}$ is the metric on a $D-2$ dimensional sphere. From the field equations, the following relation is obtained between $\tau$ and $\rho$:
\begin{equation}
\tau - \rho=(D-2)[1+2\bar{\alpha}Q][N-P] 
\label{eq27}
\end{equation}
Here, $\bar{\alpha}=(D-3)(D-4)\alpha$. $N$ and $P$ are parameters occurring in the expressions of the curvature two forms obtained from the Cartan structure equations given by: 
\begin{eqnarray}
N=-\frac{\Phi '}{r}(1-\frac{b}{r})  \label{eq28a}\\
P= \frac{1}{2r^{3}} (b'r-b) \label{eq28b}
\end{eqnarray}
This relationship between $\rho$ and $\tau$ differs from the one in GTR only due to the presence of two extra factors $(D-2)$ and $[1+2\bar{\alpha}Q]$. Now at the throat, Equation \eqref{eq27} leads to: 
\begin{equation}
\tau_{o} - \rho_{o}=(D-2)\left[1+\frac{2\bar{\alpha}}{{b^{2}}_{o}}\right][N-P]_{r=b=b_{0}}
\label{eq29}
\end{equation}
Thus, from the WEC inequality the matter near the throat will be normal or exotic if the quantity $(1+2\bar{\alpha}/{b^{2}}_{o})$ is negative or positive respectively. This leads to constraints on $\alpha$ and $b_{o}$. If $\alpha>0$ the quantity $(1+2\bar{\alpha}/{b^{2}}_{o})$ is always positive and hence the WEC is violated at the throat.For a negative value of $\alpha$, $(1+2\mid \bar{\alpha}\mid /{b^{2}}_{o})$ can be positive or negative for $b_{o}>\sqrt{2 \mid \bar{\alpha}\mid} $ or $b_{o}<\sqrt{2 \mid \bar{\alpha}\mid} $ respectively. Thus matter threading the wormhole can be normal at the throat. It is also seen that for $D>5$, exotic mater can be limited to an arbitrarily small region, in contradiction to 4D GTR. \\
It was not possible however to construct a wormhole with normal matter everywhere. It is useful to note here that from the perspective of string theory, $\alpha$ can indeed be positive. For example, in heterotic string theory, $\alpha $ is interpreted as the inverse string tension and is positive-definite. The assumption in \cite{PhysRevD.46.2464} that wormhole solutions with normal matter everywhere do not exist in EGB theory is based on the positivity of the quantity $N-P$ in equation \eqref{eq27}.  This positivity is a requirement only near the throat and is not valid for the entire space-time.  It was shown in \cite{PhysRevD.78.024005} that wormhole solutions with $\alpha >0$ can exist with normal matter everywhere with the assumption that the space-time has symmetries corresponding to the isometries of a $(D-2)$ dimensional maximally symmetric space with sectional curvature $k=-1$. Such a solution was not found however for $k=1$ and $\alpha<0$.  A counter example of this was explicitly obtained for the first time by Mehdizadeh et. al. in \cite{Mehdizadeh:2017tcf} wherein a wormhole solution with normal matter everywhere was obtained. \\
Wormhole solutions have also been obtained in dilatonic EGB (EGBd) Theory without the need of exotic matter. This theory has in addition to $R$, the coupling of a scalar field known as the dilaton field to the quadratic Gauss-Bonnet term. The action has the form:
\begin{equation}
S=\int d^{4}x \sqrt{-g}\left[R-\frac{1}{2}\partial_{\mu} \phi \partial^{\mu} \phi+\alpha e^{-\gamma \phi}{R^{2}}_{GB}\right]
\label{eq30}
\end{equation}
Where $\gamma$ is the coupling parameter. Various solutions to the field equations of EGBd theory were explored in \cite{ PhysRevD.54.5049}. Blackholes and cosmological solutions were obtained along with a special solution that necessarily exhibited properties of a wormhole space-time but with a pathology of the $g_{rr}$ term. This was avoided with a suitable coordinate transformation in \cite{ PhysRevLett.107.271101}, and it was shown that the solution indeed represented a wormhole. Such a wormhole in EGBd theory can be constructed in four dimensions without the need of any form of exotic matter. A spherically symmetric, static wormhole in this theory is given by the line element:
\begin{equation}
ds^{2}=-e^{2\nu(l)}dt^{2}+f(l)dl^{2}+(l^{2}+{r_{o}}^{2})d\Omega^{2}
\label{eq31}
\end{equation}
Here, the Schwarzschild coordinate $r$ has been redefined in terms of a new coordinate $l$ as $r=l^{2}+{r_{o}}^{2}$, $r_{o}$ being the throat radius. The metric functions as well as the dilaton field are finite at the asymptotic limit in this form. Also, the curvature invariants including ${R^{2}}_{GB}$ remain finite at the throat $l=0$ indicating the absence of a singularity. It was shown in \cite{Kanti:2011yv} that this metric represents a stable static, spherically symmetric wormhole solution that does not require exotic matter everywhere in 4D. However, the NEC is violated in regions near the throat and is satisfied asymptotically with a positive dilaton field. The violation of the WEC is attributed to the effective EMT generated by the Gauss-Bonnet term. Wormhole solutions in EGB Theory have also been explored in \cite{PhysRevD.76.064038, Dotti:2006cp}. Thin shell wormhole solutions in EGB theory that under suitable parameterizations can be supported by non-exotic matter sources have been obtained in \cite{Richarte:2007zz}.
\subsection{Braneworld Scenario \& Kanno-Soda effective theory}
\label{kset}
\begin{figure}
  \begin{center}
    \includegraphics[width=\columnwidth]{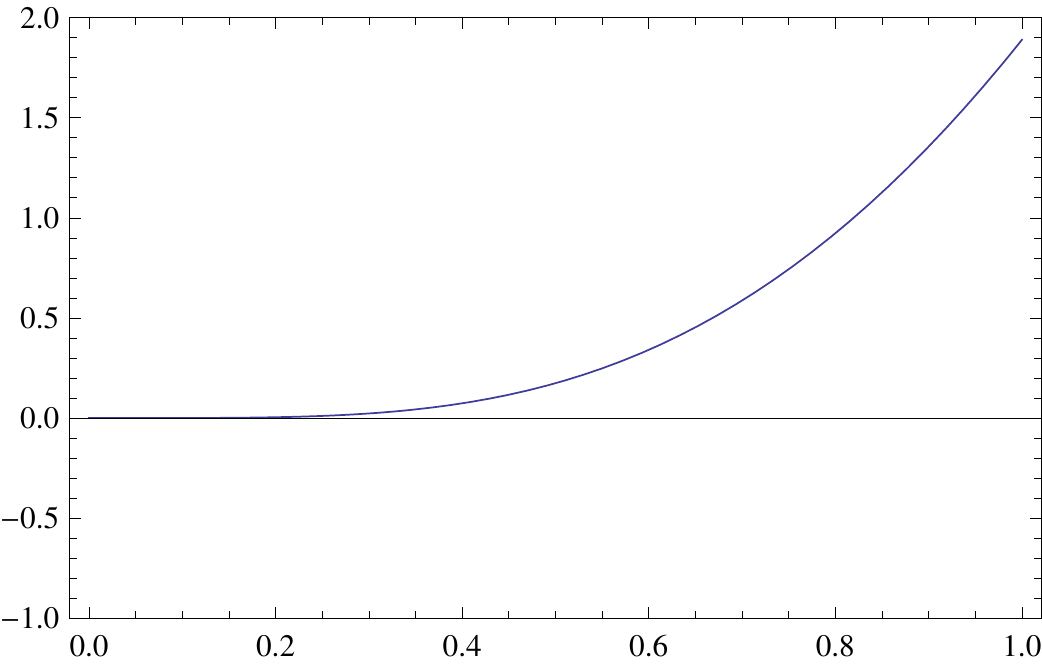}
\caption{Plot of $\rho+\tau$ vs $\mathtt{x}$ from  \cite{Kar:2015lma}}
\label{figure2}
  \end{center}
  \end{figure}

As discussed briefly in section \ref{mtged}, the braneworld scenario proposes that the observable universe is a (3+1) dimensional brane embedded in a higher dimensional bulk. The field equations in this scenario can be described on the brane by modified Einstein’s field equations of five dimensional gravity with the help of Gauss-Codazzi equations \cite{Shiromizu:1999wj}. The reduced field equations when there is no on brane matter is given by:
\begin{equation}
G_{\mu \nu} = -E_{\mu \nu}
\label{eq}
\end{equation}
Here, $E_{\mu \nu}$ is called the traceless tidal energy momentum tensor and it connects gravity on the brane with the bulk geometry. This tensor is actually the projection of the five dimensional Weyl tensor on the brane. Because of its geometric origins, $E_{\mu \nu}$ need not respect the energy conditions that are imposed upon the energy momentum tensor of ordinary matter.  Therefore in the braneworld scenario, $E_{\mu \nu}$ can serve as a matter source supporting wormhole geometries. This possibility of wormholes in a braneworld was first explored by Bronnikov and Kim \cite{PhysRevD.67.064027}. Static, spherically symmetric wormhole solutions were found for $R=0$, $R$ being the four dimensional Ricci tensor. The $R=0$ equation is an immediate consequence of the fact that $E_{\mu \nu}$ is traceless.  Lobo \cite{PhysRevD.75.064027} obtained a general class of wormhole solutions in the braneworld scenario from the context of a braneworld observer when $R \neq 0)$. Static, spherically symmetric wormhole solutions were found to exist in which the on brane matter energy momentum tensor satisfied the NEC. The violations were attributed to the effective energy momentum tensor. It was also shown that in addition to non local effects of the Weyl tensor on the bulk, as was the case in Bronnikov and Kim’s work, local high-energy bulk effects could leave a NEC violating signature on the brane. Thus, traversable wormhole geometries could arise naturally in the braneworld scenario. This approach to finding braneworld wormhole solutions is advantageous in the sense that one can consider the zero Weyl tensor case and incorporate local bulk effects by generalizing the energy momentum tensor to incorporate an anisotropic pressure contribution to the brane as was shown in \cite{PhysRevD.75.064027}. Asymptotically flat and static, spherically symmetric wormhole solutions were found in the context of the Randall-Sundrum model by Parsaei and Riazi \cite{PhysRevD.91.024015}. The work explored the consequences of the conservation of the energy momentum tensor and the traceless property of the projection of the Weyl tensor on the brane. The solutions were found to satisfy the NEC for on brane observers.
\\
An effective scalar tensor theory of gravitation in the context of the warped braneworld picture of Randall and Sundrum was developed by Kanno and Soda \cite{PhysRevD.66.083506, PhysRevD.67.104011}. The field equations for this theory are given by:
\begin{equation}
\begin{split}
\label{eq32}
G_{\mu \nu}  =&  \frac{\bar{\kappa}^{2}}{l \Psi} {T^{b}}_{\mu \nu} + \frac{\bar{\kappa}^{2}(1+\Psi)}{l\phi}{T^{a}}_{\mu \nu}\\
&+ \frac{1}{\Psi} \left(\nabla_{\mu} \nabla_{\nu} \Psi  - g_{\mu \nu} \nabla^{\alpha} \nabla_{\alpha} \Psi \right) \\ &-\frac{3}{2\Psi(1+\Psi)}\left(\nabla_{\mu} \Psi \nabla_{\nu} \Psi - \frac{1}{2}g_{\mu \nu} \nabla^{\alpha} \Psi \nabla_{\alpha} \Psi \right)
\end{split}
\end{equation}
Here, $\bar{\kappa}$ is the five dimensional gravitational coupling constant. ${T^{b}}_{\mu \nu}$ and ${T^{a}}_{\mu \nu}$ are the EMTs on the visible brane and Planck brane respectively. $\Psi$ is the \textit{radion field} which is a measure of the distance between the two branes. One can see that Brans-Dickie \cite{PhysRev.124.925} scalar tensor theory is obtained  if ${T^{a}}_{\mu \nu}=0$; the coupling parameter being $\omega(\Psi)=-\frac{3}{2\Psi(1+\Psi)}$. The scalar radion field $\Psi$ is dependent on the brane coordinates `$\mathtt{x}$' and is a measure of the distance between the two branes. It is given by: 
\begin{equation}
\Psi(\mathtt{x})=e^{2 \frac{d\mathtt{x}}{l}}-1
\label{eq33}
\end{equation} 
The theory, in the context of the RS model assumes a five dimensional bulk with a warped extra dimensional brane and two 3-branes at $y=0$ and $y=l$ respectively. $d\mathtt{x}$ in Equation \eqref{eq33} is the proper distance between the branes given by $d\mathtt{x}={\int_{0}}^{l} e^{\phi(x)}dy$. Ricci flat traversable wormhole space-time has been obtained in the KS theory by Kar et.al. \cite{Kar:2015lma} in the Jordan frame (where there is a coupling of the scalar radion field with the Ricci scalar). From Equation \eqref{eq32}, it can be seen that the effective field equations can be expressed as:
\begin{equation}
G_{\mu \nu}= \frac{\bar{\kappa}^{2}}{l\Psi}{T^{b}}_{\mu \nu}+\frac{1}{\Psi}{T^{\Psi}}_{\mu \nu}
\label{eq34}
\end{equation}
${T^{b}}_{\mu \nu}$ being the stress energy on the visible brane and ${T^{\Psi}}_{\mu \nu}$ is the stress energy on the Planck brane. In the background of a self dual spherically symmetric, static metric in isotropic coordinates of the form \cite{PhysRevD.65.064004}:
\begin{equation}
ds^{2}=-\left(\kappa+\lambda \frac{1-\frac{m}{2r}}{1+\frac{m}{2r}}\right)^{2} dt^{2}+ \frac{dr}{1-\frac{m}{2r}}+r^{2}(d\theta^{2}+sin^{2} \theta d\varphi^{2})
\label{eq35}
\end{equation}
Energy condition inequalities can be constructed from the field equations and plots of the LHS of energy condition inequalities obtained from the field equations vs. $\mathtt{x}$ are plotted to observe the behavior of the energy condition inequalities at the throat \cite{Kar:2015lma}. Here, $x=m/2r$ and the wormhole throat occurs at $r=m/2$. Hence $x=[0,1]$ with the throat being at $x=1$ and infinity at $x=0$. In order for the radion field $\Psi$ to be a stable one, it is assumed that it is never zero, and that it does not diverge into infinity at finite values of the brane coordinates. With these constraints on the radion field and some suitable choices of metric parameters, the plots of energy condition inequalities vs. the brane coordinate are obtained. The plot for $\rho+\tau$ vs. $x$ is shown in Figure \ref{figure2}. It can be seen that as $x\rightarrow 1$, $\rho + \tau$ becomes positive; implying the NEC is satisfied at the throat. The matter on the visible brane satisfies the NEC and the violations are attributed to the effective radion stress energy. The radion field is an extra dimensional entity which generates an effective geometric stress energy that allows for the satisfaction of the NEC on the visible brane. A more detailed analysis of the energy condition violations and traversabilty criteria for $R=0$ wormhole space-times in KS Theory has been done in \cite{PhysRevD.94.024011} with calculations of gravitational redshift and circular orbits.
Both stable and unstable circular orbits were found to exist and the wormhole solution is traversable for values of the metric parameters satisfying the WEC. Traversable wormholes in the braneworld scenario have also been explored in \cite{PhysRevD.91.024015,  PhysRevD.75.064027, PhysRevD.67.064027}.
\subsection{Lovelock Gravity}
\label{llg}
\begin{figure}
  \begin{center}
    \includegraphics[width=\columnwidth]{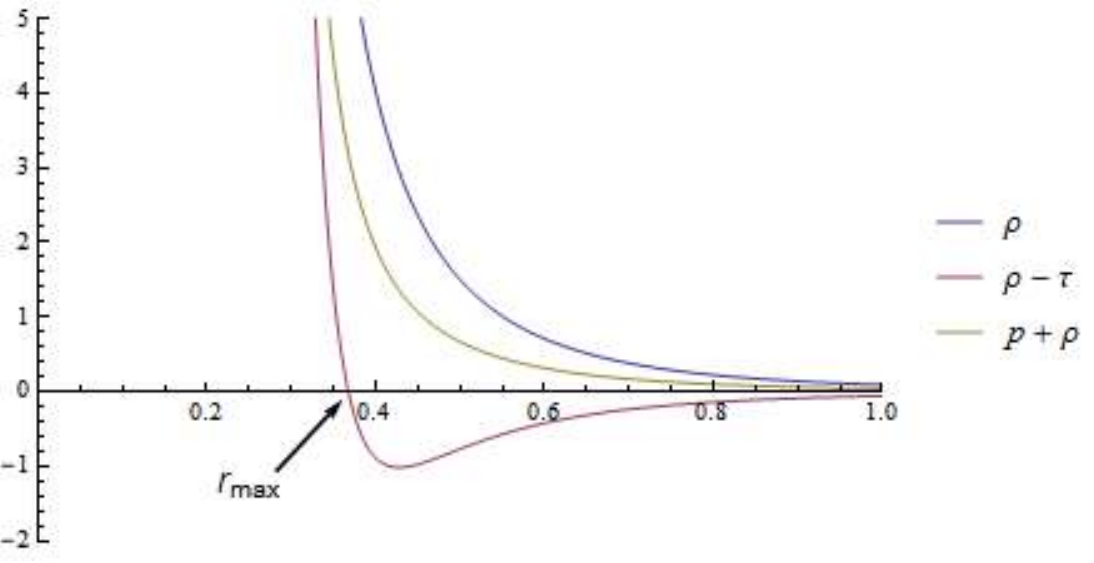}
\caption{Plot of $\rho-\tau$, $\rho+p$, and $\rho$ vs $r$}
\label{figure3}
  \end{center}
  \end{figure}

Lovelock gravity \cite{doi:10.1063/1.1665613, Lovelock1970} is the most general classical metric theory of gravitation in arbitrary number of space-time dimensions. The field equations in Lovelock gravity (upto third order and without a cosmological constant) is given by \cite{MULLERHOISSEN1985106}:
\begin{equation}
\label{eq36}
G^{(1)}_{\mu \nu}+{\sum^{3}}_{p=2} {\alpha '}_{p} \left({H^{p}}_{\mu \nu}-\frac{1}{2} g_{\mu \nu} \mathcal{L}^{p} \right)={\kappa_{n}}^{2} T_{\mu \nu}
\end{equation}
Here, $\alpha '_{p}$s are the Lovelock coefficients, $G^{(1)}_{\mu \nu}$ is the Einstein tensor and the form of ${H^{p}}_{\mu \nu}$ depends on which order of the theory is under consideration. $\mathcal{L}^{2}$ is nothing but the Gauss-Bonnet term ${R^{2}}_{GB}$ and hence it is interesting to note that the EGB theory is a generalizaion of the second order Lovelock Theory. A detailed overview of Lovelock gravity can be found in \cite{MULLERHOISSEN1985106}. The possibility of obtaining wormhole solutions in Lovelock gravity was first explored in \cite{MenaMarugan:1991ea, PhysRevD.50.3787}. Wormhole solutions in the Lovelock gravity theory have a throat size that is constrained by the Lovelock coefficients $\alpha'_{p}$, the dimensionality of the space-time and also on the shape function $\Phi(r)$. It was shown in \cite{PhysRevD.79.064010} that in third order Lovelock gravity, the geometry maybe threaded by normal matter that is confined to a region ranging from the wormhole throat to some maximum value of $r$ which is dependent on the Lovelock coefficients and the shape function. This result is of significance because it enlarges the region that can contain non-exotic matter as the source compared to second order Lovelock gravity. The field equation is given by Equation \eqref{eq36} with the third order Lovelock lagrangian \cite{MULLERHOISSEN1985106, PhysRevD.79.064010}. The Morris-Thorne background metric in $n$ dimensions is given by:
\begin{equation}
ds^{2}=-e^{2\Phi(r)}dt^{2}+\frac{dr^{2}}{1-\frac{b(r)}{r}}+r^{2}d\theta_{1}^{2}+{\sum_{i=2}}^{j=1} {\prod_{j-1}}^{i-1} sin^{2} \theta_{j} d{\theta_{i}}^{2}
\label{eq37}
\end{equation}
Field equations are derived in this Morris-Thorne background and the following expressions are obtained for checking the energy condition inequalities:
\begin{equation}
\begin{split}
\rho (r) =\frac{(n-2)}{2r^{2}}\bigl\{-\left( 1+\frac{2\alpha
_{2}b}{r^{3}}+\frac{3\alpha _{3}b^{2}}{r^{6}}\right) \frac{(b-rb^{\prime })}{%
r}  \\
 +\frac{b}{r}\left[ (n-3)+(n-5)\frac{\alpha _{2}b}{r^{3}}+(n-7)%
\frac{\alpha _{3}b^{2}}{r^{6}}\right] \bigr\}
\label{eq038}
\end{split}
\end{equation}
\begin{equation}
\rho-\tau=\frac{(n-2)}{2r^{3}}(b-rb')\left(1+\frac{2\alpha_{2} b}{r^{3}}+\frac{3a_{3} b^{2}}{r^{6}}\right) 
\label{eq38}
\end{equation}
\begin{eqnarray}
\label{eq39}
\rho+p & = & -\frac{b-rb'}{2r^{3}}\left(1+\frac{6\alpha_{2} b}{r^{3}}+\frac{15a_{3} b^{2}}{r^{6}}\right)+ \\ \nonumber &&  \frac{b}{r^{3}} \left[(n-3)+(n-5)\frac{2\alpha_{2} b}{r^{3}}+(n-7)\frac{3a_{3} b^{2}}{r^{6}}\right]
\end{eqnarray}
The equations are written in a system of units such that ${\kappa_{n}}^{2}=1$ and $\alpha_{2} \equiv (n-3)(n-4) \alpha'_{2}$; $\alpha_{3}\equiv (n-3)...(n-6) \alpha'_{3}$. A positivity of the above equations ensures that the WEC is respected implying the matter is non-exotic. Three types of shape functions:
\begin{itemize}
\item Power Law: $b(r)={r_{0}}^{m}/r^{m-1}$
\item Logarithmic: $b(r)=r\ln r_{o}/ \ln r $
\item Hyperbolic: $b(r)=r_{o}\tanh (r)/ \tanh (r_{o})$
\end{itemize}
were considered and the constraints on the Lovelock coefficients and throat radius were checked for non negative values of Equations \eqref{eq38} and \eqref{eq39}. Equation \eqref{eq39} is positive for the three types of shape functions considered in the following manner:
\begin{itemize}
\item Power law shape function: Positive for $r>r_{o}$; provided $r_{o}>r_{c}$. $r_{c}$ is a special constraint on the value of $r_{o}$ which is set by a set of constraint equations arising from the field equations along with Equations \eqref{eq38} and \eqref{eq39}. 
\item Logarithmic shape function: Positive for $r_{o}>1$; provided $r_{o}\geq \hat{r}_{c}$. $\hat{r}_{c}$ is a special constraint on the value of $r_{o}$ which is set by another set of constraint equations arising from the field equations along with Equations \eqref{eq38} and \eqref{eq39}.
\item Hyperbolic shape function: Positive for $r_{o}>\tilde{r}_{c}$, $\tilde{r}_{c}$ being another constraint arising as in the previous two cases.
\end{itemize}
The flaring out condition, $b-rb'>0$ implies from Equation \eqref{eq38} that the positivity of $\rho-\tau$ requires:
\begin{equation}
1+\frac{2\alpha_{2} b}{r^{3}}+\frac{3a_{3} b^{2}}{r^{6}}<0
\label{eq40}
\end{equation}
It is seen from that for positive values of the Lovelock coefficients, Equation \eqref{eq40} is not respected implying violation of the WEC. If either or both of the Lovelock coefficients are negative, Equation \eqref{eq40} is satisfied in the vicinity of the throat for all three types of the shape function considered provided $r_{min}<r_{o}<r_{max}$ where $r_{min}$ and $r_{max}$ are dependent on the Lovelock coefficients and the shape function via the equation:
\begin{equation}
r^{6} + 2 \alpha_{2} r^{3} b(r)+3\alpha_{3}b^{2}(r)=0
\label{eq41}
\end{equation}
The value of $r_{max}$ for the power law shape function reads:
\begin{equation}
r_{max}=\left(\frac{r_{+}}{{r_{0}}}\right)^{2/(m+2)}r_{o}; r_{+}=\left(-\alpha_{2}-\sqrt{\alpha_{2}^{2}+3\alpha_{3}}\right)^{\frac{1}{2}}
\label{eq42}
\end{equation}
This implies that the matter threading the geometry cannot be normal everywhere and is instead bounded in a region $r_{min}<r_{o}<r_{max}$ near the throat. From the above discussions, it can be seen that in third order Lovelock gravity with suitable choice of the metric parameters, the positivity demand of $\rho+p$ imposes a lower bound on the value of $r_{o}$ which is in contrast to Einstein and Gauss-Bonnet gravity. Also, the region containing normal matter is larger than that of EGB gravity. The third order Lovelock term with negative Lovelock coefficients increases the throat radius. Equation \eqref{eq40} can be generalized for $n^{th}$ order Lovelock gravity as:
\begin{equation}
1+{\sum^{[n-1]/2}}_{p=2} p \alpha_{p} \left(\frac{b}{r^{3}}\right)^{p-1} <0
\label{eq43}
\end{equation}
Equation \eqref{eq43} is satisfied upto $r_{max}<\infty$. Thus, as higher order Lovelock terms are considered in the theory with negative coupling coefficients, the throat radius grows but one cannot have normal matter everywhere for a traversable wormhole in Lovelock gravity in the Morris-Thorne background. A plot of the energy condition inequalities vs. $r$ has been shown in \ref{figure3}. Traversable wormholes in Lovelock gravity that under suitable parametrizations satisfy the energy conditions, in the presence of a cosmological constant have been obtained in \cite{Zangeneh:2015jda}. It was found that both asymptotically flat and non flat solutions exist and that a limited spherically symmetric traversable wormhole with normal matter in a 4-dimensional space-time can exist for a negative cosmological constant.
\section{Discussion and conclusion}
\label{cnd}
In this article, we have started by introducing the traversable wormhole solution of Morris and Thorne and we have also discussed some traversable wormhole solutions in modified gravity theories that respect the energy conditions. We then reviewed in detail traversable wormhole solutions in higher dimensional gravity theories with an in depth look at the mater sources threading such geometries. It is seen that in higher dimensional gravity theories, with suitable parametizations of the metric functions, the energy conditions can be satisfied by the four dimensional energy momentum tensor of the matter sources while there maybe violations of the generalized energy conditions that can be attributed to the higher order curvature terms that generate effective energy momentum tensors having an exotic behaviour. It was also seen that the coupling constants mediating the coupling of the higher order curvature terms or scalar fields in various higher dimensional theories to the four dimensional gravity sector of the field equations also require suitable parametrizations to make the four dimensional energy momentum tensor respect the energy condition inequalities. They may also impose constraints on the size of the wormholes. Aside from these well cited examples, other traversable wormhole solutions in higher dimensional theories have been studied in the context of string theory \cite{PhysRevD.44.3802, PhysRevD.44.3802, GIDDINGS198946}; superstring theory \cite{GIBBONS199637} among others, close examinations of the matter sources therein is an open and yet uncomprehensively dealt with research problem as is evident from the review of literature carried out during this work. Moreover, the solutions discussed herein are spherically symmetric and static or non static. There also exists axisymmetric, cylindrically symmetric and Eucledian traversable wormhole solutions in various higher dimensional gravity theories respecting some or all of the energy conditions that have not been taken into account here and are the subject of a review to be carried out in the future.

\printbibliography
\end{document}